\documentclass[]{an}
\usepackage{graphicx}
\usepackage{times}
\Volume{00}              
\Year{0000}              
\Month{00}               
\Pagespan{000}{000}      

\begin{document}
\lhead[\thepage]{I.~Negueruela et al.: Oe stars}
\rhead[Astron. Nachr./AN~{\bf XXX} (200X) X]{\thepage}
\headnote{Astron. Nachr./AN {\bf 32X} (200X) X, XXX--XXX}

\title{On the class of Oe stars\fnmsep\thanks{Based on
observations obtained at the Isaac Newton Telescope (La Palma,
Spain).}}

\author{I.~Negueruela\inst{1,2}, I.~A.~Steele\inst{3} \& G.~Bernabeu\inst{1} }
\institute{Dpto. de F\'{\i}sica, Ing. de Sistemas y Teor\'{\i}a de
la Se\~{n}al, Universidad de Alicante, Apdo. 99, E03080 Alicante,
Spain
\and Guest investigator of the UK Astronomy Data Centre
\and
Astrophysics Research Institute, Liverpool John Moores
University, CH41 1LD, UK} 
\date{Received {date will be inserted by the editor}; 
accepted {date will be inserted by the editor}} 

\abstract{We present high-quality spectra of the majority of stars
that have been classified as Oe and find that their published spectral
types are generally too early, most likely due to infilling of
\ion{He}{i} lines. As a matter of fact, all stars classified as Oe
actually fall inside the range O9--B0 with the important
exception of HD~155806 (O7.5\,III) and perhaps HD~39680 (difficult to
classify, but likely O8.5\,V). Observations of a sample of objects
with published spectral types in the O9--B0 range previously
classified as peculiar or emission-line stars fail to reveal any new
Oe star with spectral type earlier than O9.5. Most objects classified
as peculiar in ``classical'' 
literature show signs of binarity in our spectra, but no spectral
anomalies. We conclude that there is likely a real decline in the fraction 
of Be stars for spectral types earlier than B0, not
due to observational bias. The few Oe stars with spectral types
earlier than O9.5 deserve detailed investigation in order to provide
constraints on the physical reasons of the Be phenomenon. 
\keywords{stars: emission-line, Be --  stars: evolution --
early-type}  
}
\correspondence{ignacio@dfists.ua.es}

\maketitle

\section{Introduction}

For a very long time, it has been obvious that there is a strong
correlation between fast rotation and the Be phenomenon. Though the
exact mechanism behind the Be phenomenon is not known (see Porter \&
Rivinius 2003 for a recent review), it is a well-known fact that Be
stars rotate fast (cf. Townsend et al. 2004). Fast rotation plays a
fundamental role in the evolution of massive stars (Maeder \& Meynet
2000), as the initial rotational velocity determines the main-sequence
lifetime of a star, its post-main-sequence evolution and likely even
the nature of its final remnant after supernova explosion (Meynet \&
Maeder 2000, 2003). It is naturally tempting to think that
understanding of the connection between fast rotation and the Be
phenomenon may further our knowledge about the effects of rotation on
the whole set of massive stars.

Such hope, however, seems to have been consistently frustrated  by our
profound lack of knowledge of the ultimate causes of the Be
phenomenon. As a matter of fact, we are 
still far from even knowing if Be characteristics, which are seen in a
broad variety of stars, are always due to the same physical causes,
even if we restrict ourselves to ``classical'' Be stars, i.e., fast
rotating moderately massive stars in which the emission lines are 
produced in a circumstellar disk of material expelled from the
photosphere (cf. Porter \& Rivinius 2003). 

The main reasons (or perhaps excuses) for our ignorance are the very
complexity of the Be phenomenon and the vast range of stellar
parameters over which it occurs: a B0\,III star is a very different
beast from a B9\,V star on all accounts, in spite of which, most authors will
still expect physical mechanisms behind Be characteristics in one and
the other to be the same. Indeed, the completely
arbitrary (for Nature) fact that the Be phenomenon is confined to the
vicinity of the B spectral type (whose definition, though not 
arbitrary, certainly does not respond to any fundamental physical
reason), makes us forget that the Be phenomenon spans both sides of
the divide (rather more meaningful in physical terms) between
intermediate-mass and massive stars.

The relevance of this division comes from the fact that massive stars
have self-initiating radiative winds and the presence of these winds
may have some effect on whatever mechanisms cause the Be
phenomenon. As the Be phenomenon tends to become {\em very} rare among
O-type stars, some thought should be given to the physical causes of
this scarcity of Oe stars. Naively, one could think that a fast
radiative wind will exert a force on any circumstellar material that
will be sufficient to sweep it away and prevent it from forming a
disk. Though radiative pressure may certainly be an important factor
to prevent the formation of a Be disk, several
authors have argued that other effects may be even more determinant in
explaining the disappearence of the Be phenomenon towards early types.
In the evolutionary model proposed by Keller et
al. (2001, thoroughly discussed in Section~\ref{sec:discusion}), based
on model calculations by Meynet \& Maeder (2000), the 
reason why the Be phenomenon does not extend into the O-range is the
very different evolution of angular momentum in fast rotators of
moderate and high mass.

In order to gain some understanding of the reasons why the Be
 phenomenon is restricted to the B-type range, an investigation has
 been conducted in search of O-type stars with Be characteristics. Our aim is
 assessing the frequency of the Be phenomenon in the O-range and
 the earliest type at which the Be phenomenon is seen. A similar
 programme was recently conducted to determine the highest luminosity
 at which the Be phenomenon was observed (Negueruela 2004, henceforth
 Paper I).

\section{Target selection and Observations}

\subsection{Oe stars}
The extension of the Be phenomenon to the O-type range was first
proposed by Conti \& Leep (1974), who defined the set of Oe stars as
those O stars showing emission in the Balmer lines, but not in
\ion{He}{ii}~4686\AA\ or any \ion{N}{iii} lines. The reason to exclude
stars with \ion{He}{ii}~4686\AA\ emission is mainly morphological, but
expected to have a physical base. \ion{He}{ii}~4686\AA\ emission in
O-type stars is
generally believed to be a signature of a strong radiative
wind. Together with the presence of selective \ion{N}{iii} emission, it is
a defining characterisitc of the Of morphology, 
frequently accompanied by the presence of H$\alpha$
emission. In most cases, these \ion{He}{ii}~4686\AA\ and H$\alpha$
lines present either P-Cygni or single-peaked shapes. Hence they
should be easy to distinguish from emission lines produced in a disk
configuration, which are typically double-peaked (though this may not
be the case for optically thick lines; see Hummel \& Hanuschik
1997). Double-peaked emission lines from \ion{He}{ii}~4686\AA\ are
seen in a few anomalous objects, classed Oef by Conti \& Leep
(1974). The underlying assumption is that emission lines in Oe stars
respond to similar physical mechanisms to those generating emission
lines in Be stars. This is supported by their very similar shapes and
the lack of an obvious break in the spectral distribution. Most stars with
Of and Oef characteristics are, on the other hand, either supergiants
or of very early spectral type, thus presenting an obvious
discontinuity with the distribution of Be stars.  

\begin{table}[h]
\caption{The list of stars given as Oe objects by Frost \& Conti
(1976) with their ``classical'' spectral types. ``Modern'' spectral
types are those derived here, except for 
X~Per, taken from Lyubimkov et al. (1997). HD~46056 nd HD~203064 have
not been observed. }
\label{tab:oes}
\begin{tabular}{lcc}\hline
Object&\multicolumn{2}{c}{Spectral Type}\\ 
& Old & Modern\\
\hline
HD~24534 (X Per)& OBe& B0\,Ve \\
HD~39680& O6\,V?[n]pe& O8.5\,Ve \\
HD~45314 & OBe& B0\,IVe \\
HD~46056& O8\,V(e)& $-$\\
HD~60848& O8\,V?pe & O9.5\,IVe\\
HD~149757 ($\zeta$~Oph)& O9\,V(e)&O9.5\,IV\\
HD~155806& O7.5\,IIIe& O7.5\,IIIe\\
HD~203064 (68~Cyg)& O8\,V(e)& $-$\\
\hline
\end{tabular}
\end{table}

Conti \& Leep (1974) proposed 7 members for the class of Oe
stars and few others
have been added since then, suggesting that the frequency of
the Be phenomenon is indeed  
much lower among O-type stars than at B0-B1, where it reaches
$10-15$\% of all stars with luminosity classes III-V (Zorec \& Briot
1997). Studies of Oe stars have been extremely scant. Frost \& Conti
(1976) obtained observations of HD~39680, HD~60848 and HD~155806. They
showed Balmer emission lines similar to those of Be stars. The first
two stars also displayed double-peaked \ion{He}{i} emission
lines. Frost \& Conti (1976) noted that \ion{He}{i}~4471\AA\
was filled in by emission and therefore the spectral types obtained
using the canonical
\ion{He}{ii}~$\lambda$4541/\ion{He}{i}~$\lambda$4471 ratio would be
earlier than really corresponded to the stars. They noted that
HD~39680, previously classified O6\,V on this account, appeared to have a
spectral type around O9. They classified HD~46056, $\zeta$~Oph and
68~Cyg as O(e), because emission was seen sporadically and only in
H$\alpha$. Their list of Oe stars is reproduced in Table~\ref{tab:oes}.

Andrillat et al. (1982) obtained red and near-infrared spectra of all
the objects in Table~\ref{tab:oes} except 68~Cyg. HD~60848 and
HD~155806 showed emission in the Paschen series, similarly to early Be
stars, while HD~39680 and $\zeta$~Oph did not.

A detailed analysis of photospheric features of X Per when it was not
in a Be phase led to its classification as B0\,V (Lyubimkov et
al. 1998). Hence, this object will not be considered as an Oe
star. HD~46056 has only been reported to show some emission infilling
in the core of H$\alpha$. Both this object and 68~Cyg have been
extensively studied, with no further references to Oe
characteristics. Moreover, Underhill \& Gilroy (1990) suspected binarity
for HD~46056. Therefore these two objects are, at best, very mild
members of the class.

\subsection{Sample}

In order to investigate the extent of the Be phenomenon into the
O-star range, we selected a sample of objects chosen according to
several criteria. First, we included all stars
classified as Oe visible from La Palma. Apart from those in
Table~\ref{tab:oes}, we ran a search of SIMBAD selecting emission-line
stars of spectral type O. However, many morphologically normal Of
stars are listed as emission-line objects and were removed from the
sample. Also, a few  candidates were rejected because SIMBAD did
not provide any reference in the literature from which the
classification as emission-line stars could stem (one such example is
\object{BD $+36\degr$4032}, studied in Paper I).

In addition we selected stars with spectral types in the O8-B0 range
that had been 
classified as ``p'' (peculiar) in the classical works of Morgan et
al. (1955, from now on M55) and Hiltner (1956, henceforth H56). 
With this, we expected to be able to select some other early Be stars
and also study the possible existence of objects with peculiar
characteristics, perhaps resembling the Oef or Of?p stars (cf. Rauw
et al. 2003; Walborn et al. 2003) at later spectral types or the
peculiar O9 star \object{BD +53$\degr$2790} (Negueruela \& Reig 2001). 
All stars for which we could find in the literature an explanation for
the ``peculiarity'', such as CNO anomalies, were excluded from the sample.

\subsection{Observations}
 
Observations were collected as part of the Isaac Newton Group
service programme at the 2.5-m Isaac Newton Telescope (INT), with a
few additional spectra obtained during an observer led run at this 
telescope in July 2003. For all the observations, the Intermediate Dispersion
Spectrograph (IDS) was used with the 235-mm camera and a
combination of settings (detailed in
Table~\ref{tab:settings}). The service observing strategy allowed
the observation 
of stars over a broad range in Right Ascension. Unfortunately it
did mean that the whole sample could not be observed. However, the
sample observed is not only a large fraction ($\sim80\%$) 
of the original sample,
but certainly large enough to be considered representative.

\begin{table}[h]
\caption{Dates of observation and configurations of the
instrumentation used. All observations have been obtained with the
INT+IDS, equipped with the 235-mm camera.}
\label{tab:settings}
\begin{tabular}{llcc}\hline
Observation & Set-up & Wavelength & Dispersion\\ 
Date& & Range & (\AA/pixel) \\
\hline
2001 Aug 27& R1200R + Tek\#5&6275--7120\AA&0.8\\
2001 Oct 7& R1200B + EEV\#10 & 3600--5250\AA&0.45\\
2002 Jan 26& R1200R + EEV\#10 & 5750--7400\AA&0.45\\
2002 May 23&R1200B+Tek\#5&3850--4710\AA&0.8\\
&&4100-4940\AA&0.8\\
&R1200R+Tek\#5&6125--6970\AA&0.8\\
2003 Jul 1-6&R900V+EEV\#13&4900--7200\AA&0.65\\
\hline
\end{tabular}
\end{table}

In order to complete the sample of Oe stars, archival observations of
\object{HD~39680} and \object{$\zeta$~Oph} 
were retrieved from the ING Archive. Spectra of \object{HD~39680} had
been obtained on the night of 1997 March 25th with the INT+IDS
equipped with the 500-mm camera, R1200B grating and Tek\#3 CCD. This
setup covers 
only $\sim 400$\AA, but three different grating angles had been used
in order to span the range $\lambda\lambda$~3750-4560\AA.
The observations of $\zeta$~Oph had been obtained on 1997 April 25th,
using the INT+IDS equipped with the 235-mm camera, H2400B grating and
the Tek\#3 CCD. Two grating angles were used to cover the range
$\lambda\lambda$~3950-4750\AA. 

\begin{table}[h]
\caption{Log of observations. All observations have been obtained with
the INT (dates of observation are specified
in Table~\ref{tab:settings}), except those from Aug 2003, which were taken
with the WHT. The spectral types derived are indicated in column
4. For emission-line objects, the measured equivalent width of
H$\alpha$ is given in column 5. The uncertainty in this measurement
stems mainly from the continuum determination and it is estimated to be
at $\sim5\%$ level. Because of this, all values have been rounded up
to the nearest integer. Note that three objects were not observed in H$\alpha$}
\label{tab:obs}
\begin{tabular}{lcccc}\hline
Object& \multicolumn{2}{c}{Observed} &Spectral& EW (H$\alpha$) \\ 
& Blue & Red & Type & (\AA) \\
\hline
BD $+59\degr$2829 &Jul 03 & Aug 01 & B0.7\,Ve &$-$42\\
BD $+61\degr$105 &Jul 03 & Aug 01 & O9\,IV &abs\\
LS I $+63\degr$94 &Jul 03 & Aug 01 & B1\,IIIe& $-$36\\
BD $+60\degr$180 &Jul 03 & Aug 01 & B2.5\,IIIe &$-$30\\
BD $+54\degr$395 &Jan 01 & Aug 01 & B0.2\,V &abs\\
BD $+61\degr$370 &Oct 01 & Aug 01 & O9\,V &abs\\
HD 12856 &Oct 01 & Aug 01 & B0.5\,III-IVe &$-$36\\
LS I $+61\degr$277 &Oct 01 & Aug 01 & B1\,Ve& $-$78\\
HD 16832 &Oct 01 & Aug 01 & O9.5\,II-III & abs \\
HD 254755 &Oct 01 & Aug 01 & O9\,V &abs\\
HD 255055 &Oct 01 & $-$ & O9.5\,V(e?) &$-$\\
HD 256035 &Oct 01 & Jan 02 & O9\,V+? &abs\\
HD 45314 &Oct 01 & Jan 02 & B0\,IVe &$-$33\\
HD 46847 &Oct 01 & $-$ & O9.7\,IV & $-$\\
HD 50891 &Oct 01 & $-$ & B1\,IIIe & $-$\\
HD 60848 &May 02 & May 02 & O9.5\,IVe &$-$6\\
HD 155806 & Aug 03 & Aug 03 & O7.5\,IIIe& $-$6\\
BD $-$08$\degr$4634 &May 02 & May 02 &BN0.2\,IIIe&$-$1\\
HD 228548 &May 02 & May 02 & B1\,Ve& $-$48\\
BD $+36\degr$4145 &May 02 & May 02 & O8.5\,V& abs \\
BD $+42\degr$3835 &May 02 & May 02 & O9.5\,II-III& abs\\
BD $+45\degr$3260 &May 02 & May 02 & O7.5\,III((f))& abs\\
LS III $+58\degr$38 &May 02 & Aug 01 & O9\,V+ & abs\\
LS III $+57\degr$88 &May 02 & Aug 01 & B0.5\,Ve&$-$75\\
HD 240234&Jul 03 & Aug 03 & O9.7\,IIIe&$-$12 \\
BD $+61\degr$2408 &May 02 & Aug 01 & B0.2\,IV+ &abs\\
HD 224599 &May 02 & Aug 01 & B0.7\,IIIe &$-$6\\
\hline
\end{tabular}
\end{table}

 In addition, 
observations of two other stars (HD~155806 and HD~240234) were
obtained on 2003 Aug 17th using the 
4.2-m William Herschel Telescope (WHT) equipped with the Intermediate
Dispersion Spectroscopic and Imaging System (ISIS). The R1200B and
R1200R gratings were used in the blue and red arms respectively, which
were suited with the EEV\#12 and the MARCONI2 CCDs. These spectra have
a higher dispersion ($\sim0.2$\AA/pixel) than the INT ones. HD~240234
had already been observed in the blue with the INT.
 
\section{Results}

Spectral classification has been carried out according to the standard
criteria outlined by Walborn (1971) by comparison to standard stars
(as specified in Paper~I) and the grid of standards of Walborn \&
Fitzpatrick (1990). We have been careful, though, to take into account
the fact that \ion{He}{i} lines are generally filled in by emission
components in Be stars (see Steele et al.~1999 for a more thorough
discussion). On the other hand, all spectral types derived are based
on the assumption that Be stars do not suffer infilling of the
\ion{He}{ii} lines. The disks of Be stars, where the lines are
produced, are rather cooler than the stellar photospheres, $T_{\rm
disk} \approx 0.5 T_{\rm eff}$ (Millar et al. 2000),
and they will not reach the temperatures needed for the production of
\ion{He}{ii} emission. Moreover, Conti \& Leep (1974) based their
definition of the Oe class on the fact that no \ion{He}{ii} emission
was present, as opposed to Of or Oef stars, where emission lines are
believed to arise due to different physical mechanisms.  We now discuss
our classification of the individual objects.

\begin{figure}[ht]
\resizebox{\hsize}{!}
{\includegraphics[bb= 50 170 485 630,angle=-90]{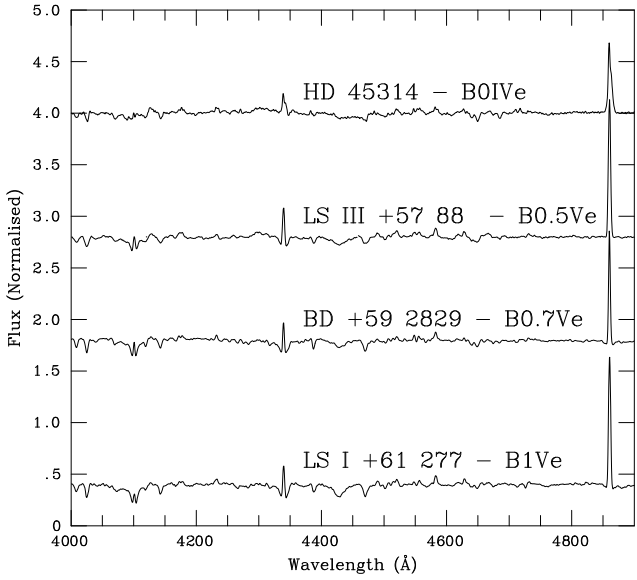}}
\caption{Four Be stars with strong metallic line emission and spectral
types close to B0. Though \object{HD~45314} is given in the original
list of Oe stars, the spectrum presented here only shows weak  
\ion{He}{ii}~4868\AA, indicating B0\,IV. The classification of all
these objects is somewhat uncertain, because of the
very strong metallic emission.}
\label{fig:bes}
\end{figure}

\subsection{BD +59$\degr$2829}
Known as a Be star since the work of Merrill \& Burwell (1933),
\object{BD~+59$\degr$2829} was classified B0\,IV?npe by M55. It lies in
the region traditionally assigned to Cas~OB5.

Our spectra show that this star is still in the Be phase, displaying
very strong narrow Balmer lines, weak \ion{He}{i} emission lines and
quite heavy metallic line emission (see Fig.~\ref{fig:bes}).
 
\ion{He}{ii}~4686\AA\ is only very weakly present, suggesting
B0.7, while the very moderate strength of
\ion{C}{iii}~4650\AA\ and the \ion{Si}{iv}~4089,4116\AA\ doublet makes it a
main sequence star. We therefore adopt B0.7\,Ve.

\subsection{BD +61$\degr$105}

Given as an emission line star in the LS catalogue
-- LS~I~+62$\degr$118, OB$^{+}$(h,le) -- \object{BD +61$\degr$105} was
classified O9\,V by M55. There is no sign of emission components in our
spectra (the blue spectrum is displayed in Fig.~\ref{fig:o9s}), but
the strength of the \ion{Si}{iv}~4089,4116\AA\ doublet and 
\ion{He}{ii}~$\lambda4686\simeq\,$\ion{C}{iii}~$\lambda$4650 support a
luminosity class IV leading us to our final classification of O9\,IV.

\subsection{LS I +63$\degr$94}

Listed as a Be star by Merrill \& Burwell (1949),
\object{LS~I~+63$\degr$94} is given as B0\,III?pe by H56. Our spectrum
displays H$\alpha$ strongly in emission and double-peaked H$\beta$ and
red \ion{He}{i} lines, however no \ion{Fe}{ii} emission lines are obvious.

\begin{figure}[ht]
\resizebox{\hsize}{!}
{\includegraphics[bb= 50 170 485 630,angle=-90]{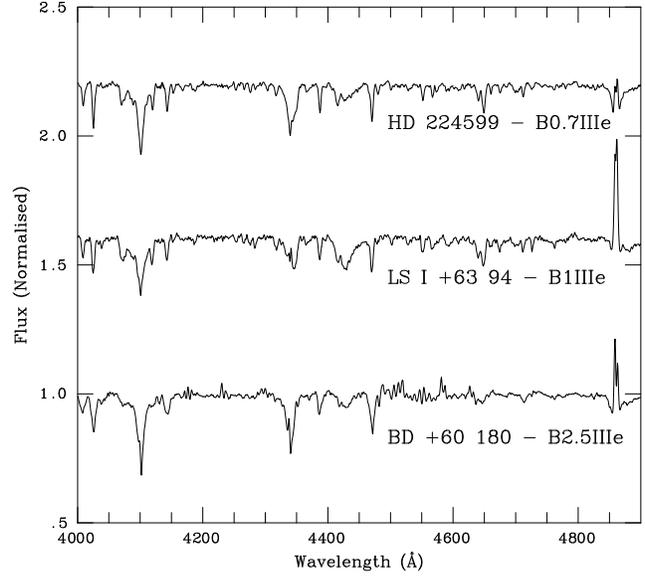}}
\caption{Three Be stars of relatively high luminosity. \object{BD
+60$\degr$180} clearly has a spectral type much later than all the
other stars in the sample. Its early previous classification might
have been caused by the very strong double-peaked metallic lines that
veil its spectrum.}
\label{fig:morebes}
\end{figure}

The blue spectrum of \object{LS~I~+63$\degr$94} is shown in
Fig.~\ref{fig:morebes}. No \ion{He}{ii} lines are visible in the
spectrum, but \ion{C}{iii}/\ion{O}{ii}~4650\AA\ is
prominent. The spectral 
type is B1\,III.
 
\subsection{BD +60$\degr$180}
Known as a Be star since the work of Merrill \& Burwell
(1933), \object{BD~+60$\degr$180} was classified B0?pe by M55. In our
spectra, it shows moderately strong double peaked H$\alpha$ and H$\beta$
emission. In the higher Balmer lines and \ion{He}{i} lines, only the
dominant blue peak is seen.
Double-peaked \ion{Fe}{ii} lines are rather strong in the green region
and heavily veil the blue spectrum (see Fig.~\ref{fig:morebes}).

The general aspect of the spectrum shows that it is considerably later
than previously assumed. The \ion{Si}{ii}~4128\AA\ doublet is
rather strong and \ion{Mg}{ii}~4481\AA\ is a prominent line,
while \ion{O}{ii}~4650\AA\ is rather weak and less prominent
than \ion{N}{ii}~4640\AA. The spectral type is around
B2.5\,IIIe.

\subsection{BD +54$\degr$395}

Classified B0\,IV?p by M55, this star lies relatively far away from the
Galactic plane ($b=-6\fdg4$) and suffers from little reddening. We
find nothing anomalous in its spectrum, which is indeed extremely
similar to the B0.2\,V standard $\tau$~Sco. Using this luminosity class and
the photometry of
H56, we find a distance modulus $DM=12.8$ which is compatible with the Perseus
Arm. 

\subsection{BD +61$\degr$370}

\begin{figure*}[ht]
\resizebox{\hsize}{!}
{\includegraphics[bb= 105 140 435 670,angle=-90]{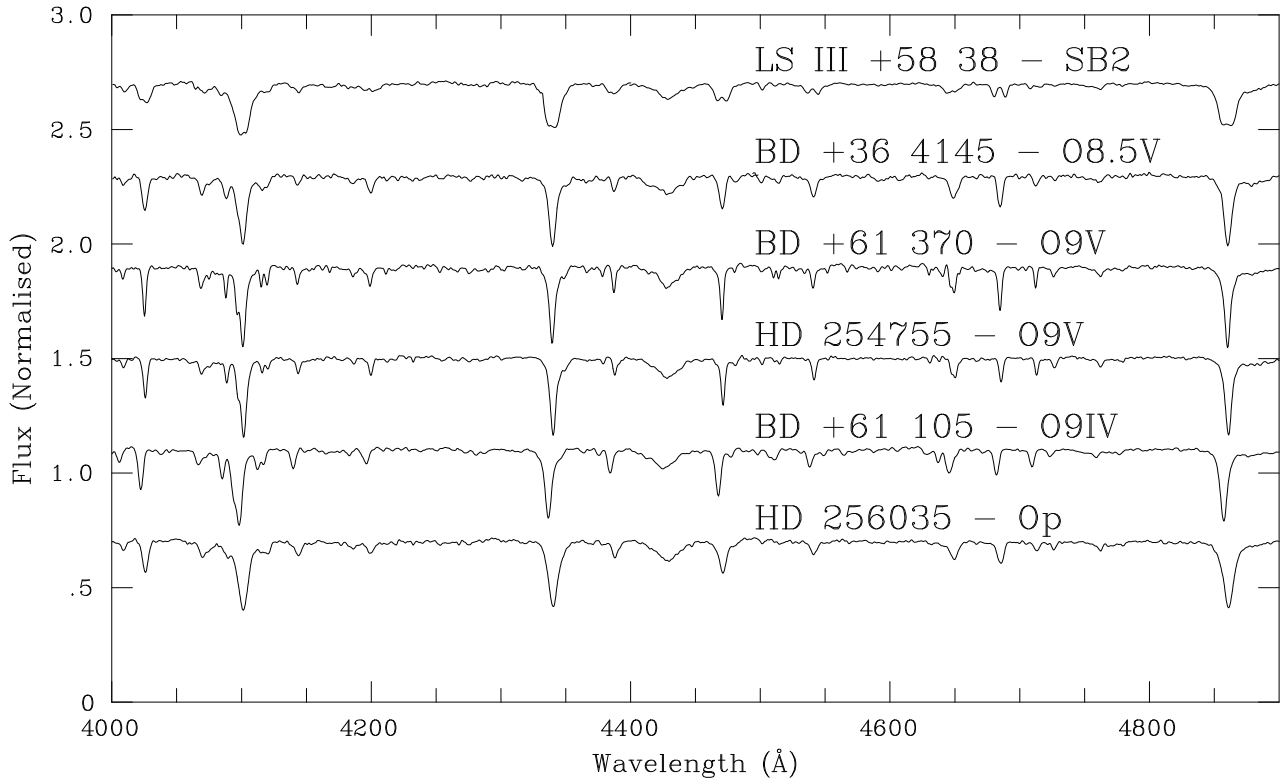}}
\caption{Spectra of sample objects with spectral types close to O9 not
displaying emission lines. \object{LS~III~+58$\degr$38} is a
double-lined spectroscopic binary with one of the components slightly
earlier than the other. \object{BD~+61$\degr$370} is a narrow-lined  
O9\,V star very similar to the MK standard \object{10~Lac} (note 
\ion{He}{ii}~$\lambda4686\:\gg\:$\ion{C}{iii}~$\lambda$4650). In    
\object{BD +61$\degr$105}, metallic lines are rather stronger and 
\ion{He}{ii}~$\lambda$4686$\:\simeq\:$\ion{C}{iii}~$\lambda$4650.
\object{HD~254755} is intermediate between the two. In
\object{HD~256035}, line ratios are anomalous, most likely because of
two different spectral components.}
\label{fig:o9s}
\end{figure*}

Even though it is noted as an emission-line star in the LS catalogue
-- LS~I~+61$\degr$256, OBce,(le) --,  BD +61$\degr$370 was simply
classified O9\,V by H56. Our spectrum, displayed in Fig.~\ref{fig:o9s},
fully confirms the classification 
and no anomalies are noted.

\subsection{\object{HD 12856} = BD +56$\degr$429}

Within the theoretical boundaries of Per OB1, \object{HD 12856} was
given as a Be star by Merrill \& Burwell (1933) and
classified B0pe by M55. Our spectrum confirms that it is still in the
Be phase, displaying double-peaked Balmer-line emission (H$\alpha$ is
single-peaked)  and moderately strong metallic emission lines. 
\ion{He}{ii}~4686\AA\ is only very weakly present, while
\ion{C}{iii}~4650\AA\ is very   
strong. The spectral type is hence close to B0.5. The luminosity class
is difficult to establish, but the strength of
\ion{C}{iii}~4650\AA\ supports III, though  
\ion{S}{iv}~4089\AA\ appears a little too weak.

Using the photometry of H56, we find $DM=12.1$ for B0.5III and
$DM=11.6$ for B0.5IV, which is close to modern determinations of the
distance to $h$ and $\chi$~Per.

\subsection{LS I +61$\degr$277}

LS I +61$\degr$277 has been known as a Be star for a very long time
(cf. Merrill \& Burwell 1933). It is believed to be a member of the
young open cluster IC~1805 and it is given as O9\,Ve in SIMBAD and
O9.5\,Ve by Shi \& Hu (1999). 

Though the heavy emission veiling makes an accurate classification
difficult, the classification as Oe star is not
tenable, based on our spectrum (shown in Fig.~\ref{fig:bes}). No
\ion{He}{ii} lines are seen and 
\ion{S}{iv}~4089\AA\ is hardly present. Considering the
weakness of  \ion{C}{iii}~4650\AA, a spectral type around
B1\,V is most likely.

\subsection{\object{HD 16832} = BD +56$\degr$703}

\begin{figure*}[ht]
\resizebox{\hsize}{!}
{\includegraphics[bb= 105 140 435 670,angle=-90]{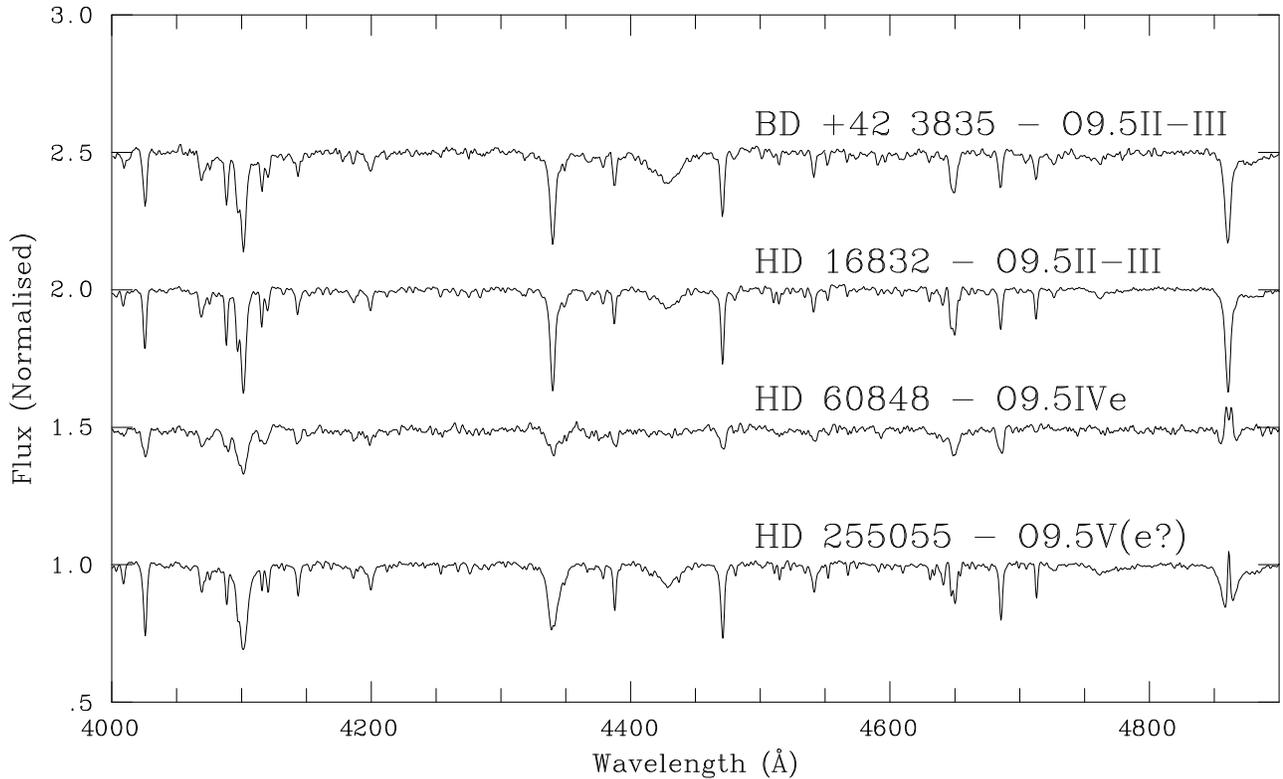}}
\caption{Spectra of sample objects with O9.5 spectral type. From the
strength of the \ion{S}{iv}~4089\AA\ line, \object{HD 16832}
and \object{BD +42$\degr$3835} are close to luminosity class II, but
\ion{He}{ii}~4686\AA\ is still a bit too
strong. \object{HD~60848}, one of the ``classical'' Oe stars is here
shown to be much later than previously thought. The condition
\ion{He}{ii}~$\lambda$4686$\:\simeq\:$\ion{C}{iii}~$\lambda$4650 makes
its luminosity class IV. \object{HD~255055} looks normal except for
the narrow emission core in H$\beta$.}
\label{fig:o95s}
\end{figure*}

\object{HD 16832} was classified B0p by M55, though later given as
O9.5\,II by Walborn (1976), who did not find any CNO anomaly in its
spectrum. Our spectrum, shown in Fig.~\ref{fig:o95s}, supports
Walborn's classification, though the 
luminosity could actually be as low as III. We find no anomalies. 
From the photometry of H56, we find a $DM=12.3$ even for O9.5III,
suggesting that -- as is generally the case of the O-type stars
assigned to Per OB1 --
\object{HD 16832} is not physically associated with $h$ and $\chi$
Per (cf. Walborn 2002 for a discussion of this issue). 

\subsection{\object{HD 39680} = BD +13$\degr$1026}
Listed as a Be star by Merrill \& Burwell (1949), \object{HD~39680} is
one of the original Oe stars in the list of Conti \& Leep
(1974). Frost \& Conti (1976) argued that the O6\,V?[n]pe var classification
given by Walborn (1973) was due to the infilling of
\ion{He}{i}~4471\AA, and commented that the spectral type of  
\object{HD~39680} based on the strength of the \ion{He}{ii} lines
should be 
close to O9. Weak wind features in its UV spectrum support a later
type (Walborn \& Panek 1984). At $b=-5\fdg9$, this object is
slightly outside the Galactic plane.  

\begin{figure*}[ht]
\resizebox{\hsize}{!}
{\includegraphics[bb= 55 175 320 670,angle=-90]{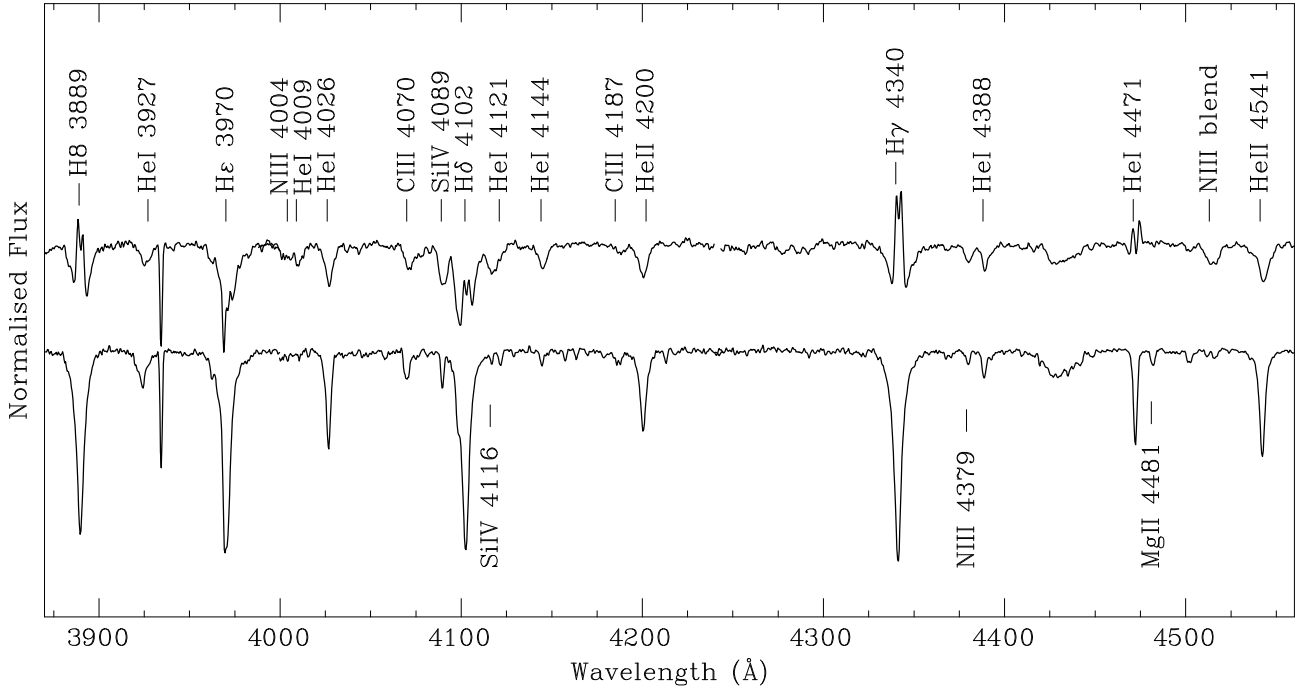}}
\caption{The spectrum of \object{HD~39680} (top) compared to the O6.5\,V
star \object{HD~42088}, observed with the same
setup. \ion{He}{i}~4471\AA\ is clearly in emission, something
seldom seen in Be stars. Very clearly \object{HD~39680} is rather
later than \object{HD~42088} and a spectral type O8.5\,V seems justified.}
\label{fig:notO6}
\end{figure*}

Our spectrum (see Fig.~\ref{fig:notO6}) fully supports the conclusions
of Frost \& Conti (1976). As a matter of fact,
\ion{He}{i}~4471\AA\ is a double-peaked emission feature,
something extremely unusual in a Be star. All the Balmer lines display
similarly double-peaked emission. An exact spectral type cannot be
derived because of the infilling of all \ion{He}{i} lines, but O8.5\,Ve
looks most likely and the star is certainly not earlier than O8, based
on the strength of the \ion{Si}{iv} lines and the presence of
\ion{He}{i}~4009\AA. All \ion{N}{iii} lines are relatively
strong, suggesting a moderate N-enhancement.

\subsection{\object{HD 254755} = BD +22$\degr$1273 }
A likely member of Gem OB1, \object{HD 254755} was classified O9\,Vp by
M55. Schild \& Berthet 
(1986) suggested that it was an N-enhanced object, but its spectrum
looks absolutely normal in this respect (see Fig.~\ref{fig:o9s}). In
the red spectrum, both 
H$\alpha$ and \ion{He}{i}~6678\AA\ are asymmetric, perhaps
suggesting binarity.

\subsection{\object{HD 255055} = BD +23$\degr$1304}

\object{HD 255055}, a likely member of Gem OB1, was classified O9\,V?p
by M55 and O9\,V(e?) by Crawford et al. (1955). The star is immersed in
diffuse nebulosity and shows very narrow H$\beta$ and H$\gamma$
emission lines (not clearly seen in the spectrum shown in
Fig.~\ref{fig:o95s}, which has been smoothed for display). As we could
not obtain an H$\alpha$ spectrum, we 
cannot decide whether the lines arise from the nebulosity or the star
itself. Otherwise, its spectrum is very similar to the O9.5\,V standard
HD~34078, except for slightly weaker \ion{C}{iii} features.

\subsection{\object{HD 256035} = BD +22$\degr$1303}

Another likely member of Gem OB1, \object{HD 256035} was classified
O9\,V?p by M55. Its spectrum, shown in Fig.~\ref{fig:o9s}, is indeed
peculiar and difficult to 
classify. All the lines are broad. The strength of the 
\ion{He}{ii} lines supports the O9 spectral type, but the \ion{Si}{iv}
lines appear too weak and the \ion{C}{iii} lines, too strong. Both
H$\alpha$ and \ion{He}{i}~6678\AA\ appear split, which,
together with the weakness of all the features in the blue spectrum,
is highly suggestive of a spectroscopic binary. 

Using the photometry of Haug (1970), we find $DM=11.0$, which is in
very good accord with the distance to Gem OB1 ($DM=10.9$; Humphreys
1978) for a single star.

\subsection{\object{HD 45314} = BD +14$\degr$1296}

A likely member of Mon OB1, \object{HD 45314} was first given as a Be
star by Merrill et al. (1925). Classified O9?pe by M55, it is one of
the objects making up the original list of Oe stars by Conti \& Leep
(1974). The spectrum of  \object{HD 45314}, shown in
Fig.~\ref{fig:bes}, displays extremely heavy 
emission veiling and few features suitable for classification are
visible. Based on the presence of \ion{He}{ii}~4200\AA\, it
must be B0 or earlier. \ion{He}{ii}~4686\AA\ is, however,
rather weaker than \ion{C}{iii}~4650\AA. 

\begin{figure}[ht]
\resizebox{\hsize}{!}
{\includegraphics[bb= 30 125 510 660,angle=0]{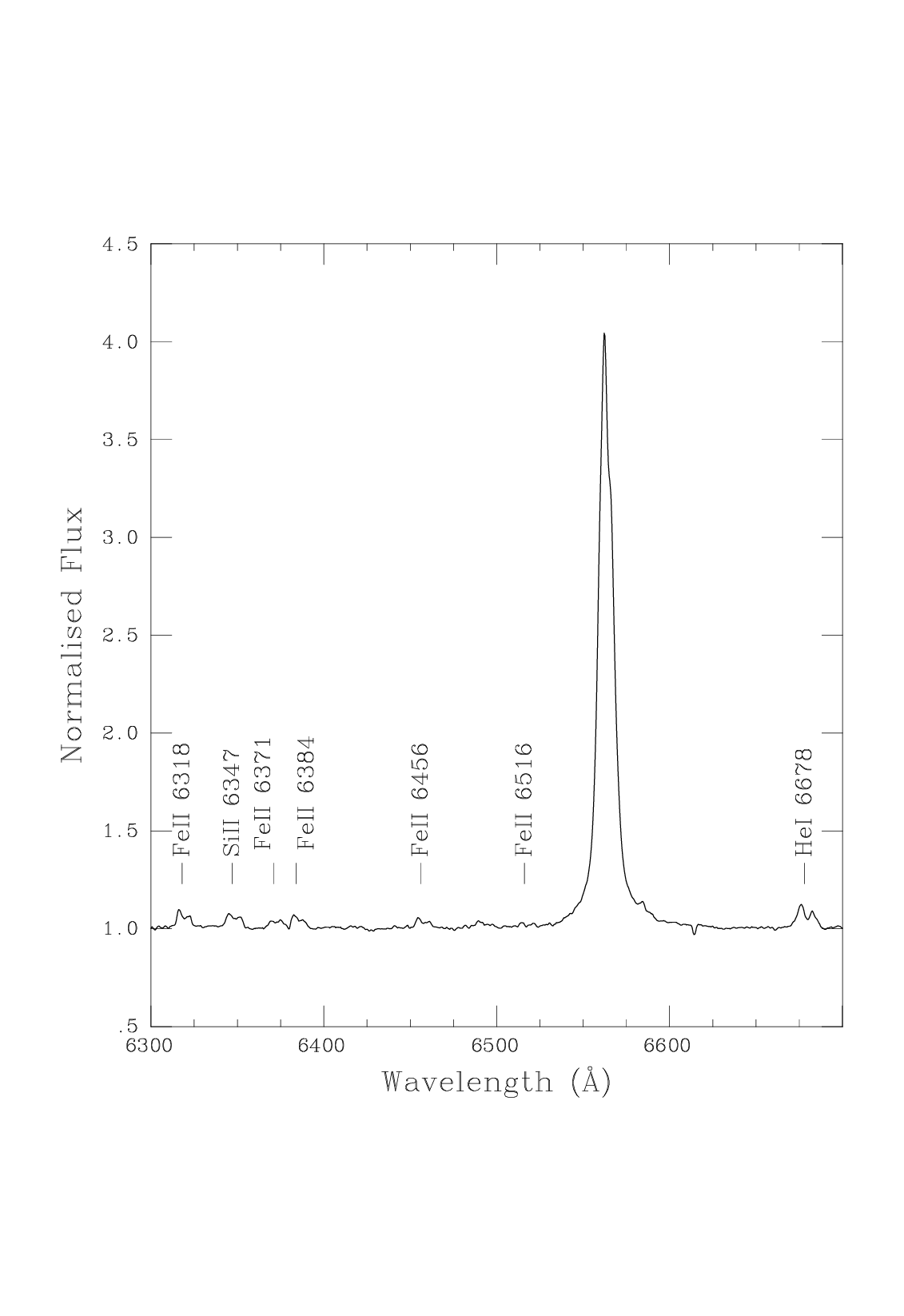}}
\caption{The spectrum of \object{HD~45314} close to H$\alpha$ is shown
as representative of stars with strong Be characteristics. H$\alpha$
is asymmetric, but the double-peaked structure is lost due to
radiative transfer effects in this optically thick line. Weak features at
$\lambda$6492\AA\ and 
$\lambda$6584\AA\ could be due to \ion{N}{ii}, though the possibility
that they are \ion{Ti}{ii} and \ion{Zr}{ii} respectively cannot be
ruled out.}
\label{fig:superred}
\end{figure}

This object is a close binary, resolved by speckle imaging with a
separation $d=0\farcs05$ by Mason et al. (1998), who do not provide
a magnitude difference between the two components. If the spectrum
observed is due to only one star, its spectral type is likely to be
B0\,IV, though O9.7\,III cannot be ruled out. In any case, it seems very
unlikely that the system contains a star earlier than O9.5.

The red spectrum of \object{HD 45314} is shown in
Fig.~\ref{fig:superred}, as it is typical of Be stars with strong
emission features. It is worth noting that this
object displayed only weak H$\alpha$ emission in 1981 February
(Andrillat et al. 1982). Such changes are typical of Be stars.  

\subsection{\object{HD 46847} = BD +02$\degr$1302}

\begin{figure}[ht]
\resizebox{\hsize}{!}
{\includegraphics[bb= 55 200 330 630,angle=-90]{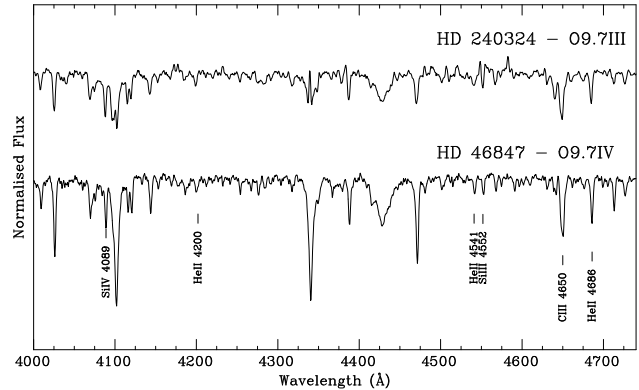}}
\caption{The spectra of two very late O stars in our sample. The
spectral type O9.7 is defined by the condition
\ion{He}{ii}~$\lambda4541 \simeq\:$\ion{Si}{iii}~$\lambda4552$. In
\object{HD~240324}, the \ion{Si}{iii} lines may be affected by
emission components, but the strength of the \ion{He}{ii} lines
guarantees that the star is earlier than B0. The strong
\ion{Si}{iv}~4089\AA\ line indicates a higher luminosity than
\object{HD~46847}.}
\label{fig:lates}
\end{figure}

A likely member of Mon OB2, \object{HD 46847} was classified B0\,III?p
by M55. According to the WDS catalogue, it is a very close visual
binary, with a second component $1.0\:$mag fainter than the
primary. There is no evidence, however, of double lines in our
spectrum, shown in Fig.~\ref{fig:lates}.

If the spectrum corresponds to a single star, the condition
\ion{He}{ii}~$\lambda4541 =\:$\ion{Si}{iii}~$\lambda4552$ defines
O9.7. As \ion{He}{ii}~4686\AA\ is weaker than
\ion{C}{iii}~4650\AA, a luminosity class IV or III is
suggested, with the moderate strength of \ion{S}{iv}~4089\AA\
supporting the former, which we therefore adopt.

\subsection{\object{HD 50891} = BD $-$03$\degr$1643}

This Be star, first listed by Merrill \& Burwell (1949), was
classified B0?pe by M55. It is still in the Be phase 
and displays moderately strong single-peaked Balmer emission
lines. The absence of any \ion{He}{ii} lines in its spectrum makes it
later than B0, while the prominent
\ion{C}{iii}/\ion{O}{ii}~4650\AA\ line means it cannot be
much later.  We therefore adopt the most likely class of B1\,IIIe.

\subsection{\object{HD 60848} = BD +17$\degr$1623}

\begin{figure}[ht]
\resizebox{\hsize}{!}
{\includegraphics[bb= 30 125 510 660,angle=0]{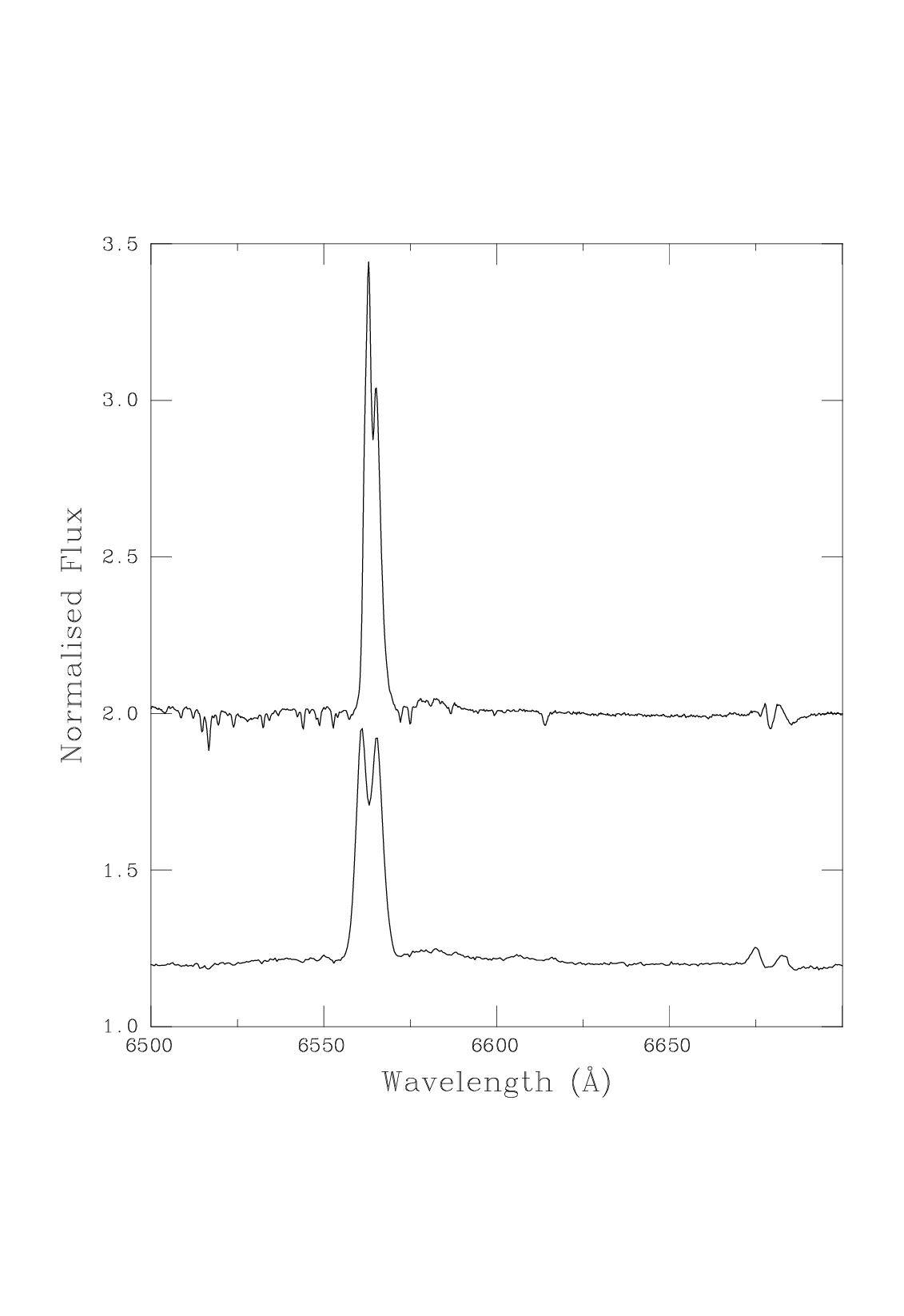}}
\caption{Red spectra of two Oe stars, \object{HD~155806} (top) and
\object{HD~60848},
showing the shape of the H$\alpha$ and \ion{He}{i}~6678\AA\
emission lines. The spectrum of \object{HD~155806} displays many weak
atmospheric features, as it has to be observed at a very high airmass
from La Palma.}
\label{fig:ored}
\end{figure}

First given as a Be star by Merrill \& Burwell (1933) and in the
original list of Oe stars, \object{HD~60848} = \object{BN~Gem} is a
well studied object with numerous references to 
double-peaked emission lines. H$\alpha$, H$\beta$ and
\ion{He}{i}~6678\AA\ display this shape in our spectra (see
Figures~\ref{fig:o95s} and~\ref{fig:ored}).

Very far away from the Galactic plane ($b=+17\fdg5$), \object{HD
60848} is indeed an interesting object. It was classified O8\,V?pevar by
M55, and hence 
considered a prototypical Oe star, but is given as B0 by Munch \&
Zirin (1961). Our spectrum, shown in Fig.~\ref{fig:o95s}, can hardly
justify the O8 
classification. \ion{He}{ii}~4686\AA\ is the only prominent
\ion{He}{ii} line, and it is not stronger than
\ion{C}{iii}~4650\AA. The \ion{Si}{iv} lines do not support a
high luminosity and the adopted spectral type is O9.5\,IVe. 

As was the case of HD~39680 (Frost \& Conti 1976), the early
classification of \object{HD 60848} must have been due to infilling of
\ion{He}{i}~4471\AA.


\subsection{\object{$\zeta$~Oph} = BD $-$10$\degr$4350}

This 2nd magnitude star has been extremely well studied in all
wavelength ranges. Classified as an O9.5\,V star by M55, it was given as
O9\,V(e) by Conti \& Leep (1974).  \object{$\zeta$~Oph} is a mild Be star
with relatively short periods of activity followed by long time-spans
of quiescence (cf. Kambe et al. 1993).

\begin{figure}[ht]
\resizebox{\hsize}{!}
{\includegraphics[bb= 55 200 330 630,angle=-90]{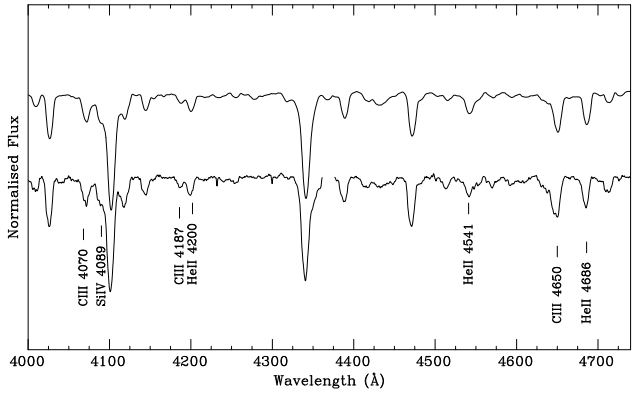}}
\caption{The spectrum of $\zeta$~Oph (bottom) is compared to that of
the MK O9.5\,V standard HD~34078 artificially spun up to a rotational
velocity $v\sin i=400\:{\rm km}\,{\rm s}^{-1}$, estimated by Herrero et
al. (1992) for $\zeta$~Oph. The metallic lines appear slightly
stronger in $\zeta$~Oph, while the \ion{He}{ii} lines are slightly
weaker, suggesting luminosity class IV.}
\label{fig:zeta}
\end{figure}

The spectrum shown in Fig.~\ref{fig:zeta} is typical of a very fast
rotator, which makes its spectral classification difficult. As
\ion{He}{ii}~$\lambda$4686$\:<\:$\ion{C}{iii}~$\lambda$4650, it
cannot be O9, unless it is of high luminosity. As there appears to be
a rather prominent \ion{O}{ii}~4640\AA\ line blended into the
\ion{C}{iii} line and the \ion{Si}{iii} triplet is still present, we
favour a spectral type O9.5\,IV, which is in good agreement with the
atmospheric parameters derived by Herrero et al. (1992).

\subsection{\object{HD 155806} = CPD $-$33$\degr$4282}

The prototypical Oe star, \object{HD 155806} was first reported as a
Be star by Merrill et al. (1925). It was classified as O8\,Ve by Hiltner
et al. (1969) and O7.5\,IIIe by Conti \& Leep (1974), and it has been
repeatedly included in different works on Be and O stars. Hanuschik et
al. (1996) reported a complex multi-peaked structure in H$\alpha$,
while Walborn (1980) reported \ion{Fe}{ii} emission lines in the
yellow region of the spectrum.

\begin{figure}[ht]
\resizebox{\hsize}{!}
{\includegraphics[bb= 55 200 330 630,angle=-90]{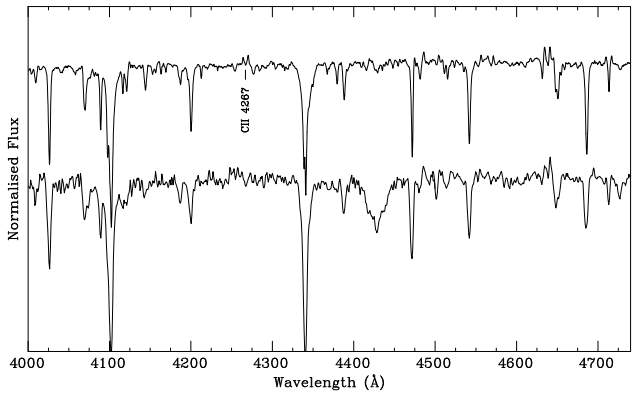}}
\caption{The spectrum of \object{HD~155806} (top) is typical of an
O7.5\,III((f)) star, except for the weak double-peaked emission lines in
\ion{He}{i}~4713\AA, \ion{C}{ii}~4267\AA\ and
\ion{Mg}{ii}~4481\AA. Note also the presence of selective
\ion{N}{iii} lines around $\lambda$4640\AA. The spectrum of the only
star of similar spectral type observed in this sample,
\object{BD~+45$\degr$3260} is included for comparison (bottom). Note
that the resolution and S/N ratio of this spectrum are rather poorer
than those of the \object{HD~155806} spectrum.}
\label{fig:theone}
\end{figure}

In our red spectrum, H$\alpha$ is an asymmetric double-peaked
structure, while \ion{He}{i}~6678\AA\ shows weak
double-peaked emission (see
Fig.~\ref{fig:ored}). In the blue, we observe no \ion{Fe}{ii}
emission lines, but \ion{C}{ii}~4267\AA\ and
\ion{Mg}{ii}~4481\AA\ show double-peaked emission (see
Fig.~\ref{fig:theone}). These 
lines are not normally seen in emission in later Be stars. In
addition, \ion{He}{i}~4713\AA\ also shows double-peaked
emission and there are weak selective \ion{N}{iii} emission lines,
typical of a mild O((f)) star. Note that the presence of these
\ion{N}{iii} lines is not consistent with the definition of Oe star
given by Conti \& Leep (1974). They are, however, morphologically
normal for a O7.5\,III spectral type. Based on the strength of
metallic features (and specifically \ion{Si}{iv}~4089\AA), we adopt
O7.5\,IIIe, though \ion{He}{ii}~4686\AA\ is rather strong and
hence the luminosity is on the low side. Indeed Walborn (1973)
classified it O7.5\,V[n]e.

\object{HD 155806} is therefore the earliest known Be star showing 
both emission lines typically seen in Be stars and others which
are not typical of Be stars of later types.

\subsection{BD $-$08$\degr$4634}

Classified O9?\,V?p by M55, \object{BD $-$08$\degr$4634} has never been
listed as an emission-line object. Our spectrum , however, reveals
weak double-peaked H$\alpha$ emission, with a deep central
reversal going below the continuum level. This shape is typical of Be
shell stars, though the low quality of our spectrum does not allow us
to ascertain whether this object is a Be shell star. 
Our
blue spectrum is not of a very high quality either, but the
photospheric lines are rather
sharp. \ion{He}{ii}~4686\AA\ is weak and the \ion{Si}{iii}
triplet is rather strong. The spectral type is likely B0.2\,III, but
Nitrogen is moderately enhanced, giving a final classification of
BN0.2IIIe.

\subsection{\object{HD 228548} = BD +39$\degr$4098}

Known as a Be star since Merrill \& Burwell (1933), \object{HD~228548}
was classified B0pe by M55. It still displays strong single-peaked
Balmer-line emission and metallic veiling. \ion{He}{i} lines show
partial infilling, except for \ion{He}{i}~4713\AA, which is in
emission. 

No \ion{He}{ii} lines are seen on the spectrum, but
\ion{C}{iii}~4650\AA\ is strong, suggesting that the spectral
type is close to B1. The \ion{Si}{iii}
triplet is completely lost in the emission forest, which would
indicate that the luminosity is low. Therefore we adopt B1\,Ve, though
a slightly earlier classification is possible.

\subsection{BD +36$\degr$4145}

Given as an emission line star in the LS catalogue --
LS~II~+37$\degr$99, OB(ce,le) --, \object{BD~+36$\degr$4145} was
classified O9\,V by M55. There is no evidence of emission in our
spectra. The classification spectrum, shown in Fig.~\ref{fig:o9s}, is
rather similar to the O9\,V standard 10~Lac, but 
the \ion{He}{ii} lines are clearly stronger, suggesting O8.5\,V.

\subsection{BD +42$\degr$3835}

BD +42$\degr$3835 was classified O9p by M55 and later given as O9\,III?
by  Garmany \& Vacca (1991). We observe no obvious peculiarities in
what appears to be a rather luminous star. Its spectrum is actually
extremely similar to that of \object{HD~16832} (see
Fig.~\ref{fig:o95s}), and we adopt the same spectral type.

\subsection{BD +45$\degr$3260}

This star, classified as O9\,V by M55, was included in this work
because of a very discrepant classification as B3\,II by Fehrenbach
(1961). Our spectrum, seen in Fig.~\ref{fig:theone}, is incompatible with both
classifications. \ion{He}{ii}~4541\AA\ is comparable to
\ion{He}{i}~4471\AA\ and selective \ion{N}{iii} lines are in
emission, but \ion{He}{ii}~4686\AA\ is relatively strong. The
spectral type is hence O7.5\,III((f)).

\subsection{LS III +58$\degr$38}

This little studied star was classified O9p by Georgelin et al. (1973)
and later given as O9\,III by Garmany \& Vacca (1991). Though the red
spectrum appears normal, in the blue spectrum all lines are clearly
double even at this moderate resolution (see Fig.~\ref{fig:o9s}). In
both spectral components \ion{He}{ii}~4686\AA\ is stronger than
\ion{C}{iii}~4650\AA\ and one of the components seems to have
rather strong
\ion{He}{ii}~4541\AA. \object{LS~III~$+58\degr$38} is hence a
double-lined spectroscopic binary. The spectral types of the
components are likely to be O8 and O9 approximately.

\subsection{LS III +57$\degr$88}

First listed as a Be star by Merrill \& Burwell (1943),
\object{LS~III~+57$\degr$88} was classified B0?\,III?pe by H56. 
It is still in a very strong Be phase, presenting very heavy metallic
emission. 

Our spectrum (see Fig.~\ref{fig:bes}) shows that
\ion{He}{ii}~4686\AA\ is weakly present, but the \ion{Si}{iv} 
lines are lost in the broad wings of H$\delta$. Though an exact
spectral type cannot be given, B0.5\,Ve is most likely.
  
\subsection{\object{HD 240234} = BD +59$\degr$2677}

Catalogued as a Be star by Merrill \& Burwell (1949), \object{HD
240234} was classified O7e by Mayer \& Mac\'ak (1973). In our red
spectrum, H$\alpha$ is rather strongly in emission and there are many
weak double-peaked emission lines corresponding to \ion{He}{i} and
\ion{Fe}{ii}. 

The blue spectrum (shown in Fig.~\ref{fig:lates}) is affected by a
moderately strong emission line forest. Both
\ion{Si}{iv}~4089\AA\ and \ion{C}{iii}~4650\AA\ are
very strong. \ion{He}{ii}~4686\AA\ is slightly weaker and
\ion{He}{ii}~4200\AA\ is clearly seen. Considering the
condition
\ion{He}{ii}~$\lambda$4541$\,\simeq\,$\ion{Si}{iii}~$\lambda$4552,
we take O9.7\,IIIe. We note that \ion{Si}{iii}~4552\AA\
could actually be suffering from blanketing by some emission line, but
the strength of the \ion{He}{ii} lines does not allow a classification
later than B0\,III, so this object is most likely an Oe star, though a
very late one.

\subsection{BD +61$\degr$2408}

Classified B0\,III?p by H56, \object{BD +61$\degr$2408} was observed by
Crampton \& Fisher (1974), 
who noted double lines in its spectrum. In our spectrum, H$\alpha$ and 
\ion{He}{i}~$\lambda\lambda$6678, 7065\AA\ are all clearly split. The
secondary component, however, is rather weak compared to the primary. 

In the blue, all three \ion{He}{ii} lines are present, but
\ion{He}{ii}~4686\AA\ is not very strong. As the \ion{Si}{iv}
lines are far too weak to justify the B0\,III classification, we adopt
B0.2\,IV. 

\subsection{\object{HD 224599} = BD +59$\degr$2801}

\object{HD 224599} lies within the traditional limits of Cas OB5 and
is listed as an emission-line star in the LS catalogue --
LS~I~+59$\degr$25, OBce,h. It
was classified  B0.5?\,V?nnp by M55. In our spectrum, H$\alpha$ is a
narrow emission line of moderate intensity, while
H$\beta$ and \ion{He}{i}~$\lambda\lambda$6678, 7065\AA\ display weak
asymmetric double-peaked emission lines.

As shown in Fig.~\ref{fig:morebes}, \ion{He}{ii}~4686\AA\ is barely
noticeable, indicating a  
spectral type B0.7. At this spectral type, the strong \ion{Si}{iii} and
\ion{C}{iii} lines clearly indicate a luminosity class III. Using the
photometry of H56 and our spectral type, we find $DM=12.4$, which is
consistent with the Perseus Arm in this direction.

\section{Discussion}
\label{sec:discusion}

\subsection{Variability and reliability}

Variability is a well known characteristic of the Be phenomenon and
indeed several surveys (e.g., Steele et al. 1999; Negueruela 2004)
have found that stars classified as emission-line objects did not
display any sign of emission at the time of the observations. It is
therefore a matter of importance to be able to rely on bibliographic
data in order to assess the likelihood that a star catalogued as an
emission-line object has indeed displayed emission lines at some
time. 

In this respect, it is suggestive, though likely not statistically
significant, to note that the 12 stars in our sample included in the
Mount Wilson Catalogue (MWC, Merrill \& Burwell 1933 and
continuations) do display emission lines in our spectra, while 3 out
of 4 objects given as emission-line stars in the LS catalogue and not
included in the MWC 
were found not to display any emission. As a matter of fact,
the only object displaying emission not in the MWC
(HD~224599) seems to
be intermittently in the Be phase, as M55 did not classify it as an
emission-line object.

Before the description of the class of B[e] stars (Conti 1976), 
it was customary to
use the Bpe designation for emission-line stars displaying forbidden
lines (cf.~Underhill \& Doazan 1982). However, among the our sample,
it appears that objects classified by M55 and H56 as 
``pe'' tend to be peculiar only in the sense that the presence of very
strong emission lines makes their spectral classification
difficult. A representative spectrum is shown in
Fig.~\ref{fig:superred}: all lines are double-peaked and correspond to
permitted transitions of singly ionised metals.
Objects marked as ``p'' but not displaying emission lines
tend to appear as spectroscopic binaries at higher resolution. No
truly ``peculiar'' object has turned up from in sample.

\subsection{The Oe stars}

The list of Oe stars presented by Conti \& Leep (1974) showed a good
spread in spectral types. However, Frost \& Conti (1976) already
cautioned that the spectral classifications were insecure and likely
to be affected by line infilling in
\ion{He}{i}~4471\AA. This has been shown by our modern
spectra to be the case for all the Oe stars catalogued, except
\object{HD~155806}. 

All the stars classified as B0pe have also received later
classifications here. This is in contrast to the result obtained by
Steele et al. (1999) for a sample of Be stars spanning the B0-B9
range. There the majority of objects, especially those of late spectral
types, were found to have earlier
spectral types than given in the literature. Both effects can be
readily explained by infilling of \ion{He}{i} lines affecting the
spectral types derived from old low-resolution spectrograms.

With the new classifications, all the known Oe stars, with the
exception of \object{HD~155806} and likely \object{HD~39860}, have spectral
types in the O9-B0 range. Their spectral types are so close to
the B spectral type that they may just be seen as the tail of the
distribution of Be stars. Only \object{HD 39680} and
\object{HD~155806} seem to display characteristics not shared by
later-type  Be stars that might be
attributed to higher temperatures (\ion{He}{i}~4471\AA\ and
\ion{C}{ii}~4267\AA\ in emission, respectively).

The number of Be stars with spectral types earlier than O9.5 is very
small. To the best of our knowledge, there are four catalogued stars
reported by several 
independent sources to be emission-line stars and have ``modern''
spectral types earlier than O9.5: the
two stars discussed in the previous paragraph, \object{HD~344863}
(O9\,III), discussed in Paper~I 
and not in a Be phase at present, and \object{HD~17520}, in
\object{IC~1848}, classified O9\,V by Conti \& Leep (1974) and O8\,V by
Walborn (1971), which entered a Be phase around 1985 (Walter 1992).
In addition, \object{BQ~Cam} and
\object{LS~437}, the optical counterparts to the X-ray pulsars
\object{V\,0332+53} and \object{X\,0726$-$260}, have spectral types O9\,V or
slightly earlier (Negueruela 1998). However, these two latter stars
are certain to have been sped up to high rotational velocities because of
extensive mass transfer, and their representativity is hence
arguable. Moreover, these two stars are much fainter than the
ones included in our (and other) sample(s) and are thus drawn from a
much larger volume. They should not be included in counts based, like
the present one, on catalogues of bright stars.   

Such scarcity is most likely not due to selection effects, though the
number of O-type stars is certainly rather low to allow good
statistics. In the recent O-type
star catalogue of Ma\'{\i}z-Apell\'aniz et al. (2004), which is
complete to $V=8$ and includes many stars fainter than this,
there are $\approx80$ stars with spectral types in the O7.5-O9
range and  luminosity classes III--V. Of these, 3 are Oe stars
(HD~344863 is not included). Though we are certainly in the area of
very small number statistics, the fraction of Oe stars appears to be
rather lower than that of Be stars. Moreover, no Oe stars are known with
spectral type earlier than O7.5 .

It seems then that observations strongly indicate a real decline in the
fraction of Be stars for spectral types earlier 
than B0. Such decline is, in principle, in agreement with arguments
presented by Meynet \& Maeder (2000; see also Keller et
al. 2001). Assuming a direct connection between fast rotation and
the Be phenomenon, these authors examine the evolution of the ratio
$\Omega/\Omega_{\rm crit}$ (surface angular velocity as a fraction of
the critical breakup velocity) during the lifetime of a star. In
their models of fast-rotating stars, objects with $M_{*}\la15M_{\sun}$
evolve in such a way that $\Omega/\Omega_{\rm crit}$ {\it increases}
during the lifetime of the star. Objects with $M_{*}\ga15M_{\sun}$
display the opposite behaviour because of angular momentum loss
associated with their winds.

Keller et al. (2001) go on to argue that the higher fraction of Be
stars to non-emission B stars observed close to the main-sequence
turn-off in some young open clusters supports the idea that most Be
stars develop their emission characteristics late in their lifetime,
as their ratio $\Omega/\Omega_{\rm crit}$ approaches unity. In this
view, the scarcity of Oe stars would be a simple consequence of the
fact that the $\Omega/\Omega_{\rm crit}$ ratio decreases for stars
with masses in the O-type range. The only Oe stars would be those that
are born with rotational velocities close to the critical value.

Though such arguments may well be qualitatively valid, it is clear
that observations do not support the details of the picture presented by Keller et al. (2001). In the latest models of rotating stars by Meynet \& Maeder
(2003), stars of $20-25M_{\sun}$ start their lives with $T_{\rm
eff}=32000-35000K$, corresponding to spectral types O9--8\,V and evolve
towards higher luminosities and lower $T_{\rm eff}$, entering
the B spectral range after 5 and 5.5 Myr respectively. At this point
they would likely have spectral types B0\,IV and B0\,III
respectively. The B0--1\,III range is therefore populated by stars of
ZAMS mass in the $20-25M_{\sun}$ range on their way to the end of their H
core-burning phase, in good agreement with the mass calibration of
Vacca et al. (1996). 

If the scenario outlined above was correct, stars of
$\approx20M_{\sun}$ would only be seen as Be stars if they happened to
be born with $\Omega$ close to the critical value, and only for a
short phase close to the ZAMS. The number of Be stars among evolved
$20M_{\sun}$ stars would be very small or zero. The observational evidence
points exactly in the opposite direction: Be stars are rather more
frequent among B0--1\,III stars than among O8--9\,V stars (cf.~Paper~I and
here). As the decrease in the number of Be stars for spectral types
earlier than O9.5 seems to generally support the scheme of Meynet \&
Maeder (2000),  possible explanations for this disagreement could be
that either 1) the relation between stellar mass
and spectral type needs to be shifted to lower masses or 2), more likely, 
the discontinuity in behaviour predicted by the models
to occur at $15M_{\sun}$ actually happens at higher masses. This would 
be a consequence of the treatment of angular momentum transport by the 
models. As a matter of fact, the more modern models by Meynet \& Maeder 
(2003) show a completely different behaviour for moderately massive
stars. Instead of decreasing through the whole MS lifetime, the
$\Omega/\Omega_{\rm crit}$ ratio for stars
in the $15-25M_{\sun}$ range actually grows quite strongly some time
before the end of the H-burning phase. This results in a peak
approaching $\Omega/\Omega_{\rm crit}=1$ during the second half of the
H-burning lifetime. This 
prediction is in much better agreement with observations, as it supports a
high Be fraction among early B-type giants.

\subsection{Outlook}
The very scarcity of Oe stars makes them extremely interesting
objects. The fact is that very few stars with spectral type around 
O8\,V display Be characteristics.  Determination of
the physical parameters of these objects may therefore give us some
clue about what ``extra'' factor turns them into emission-line stars. 

In any case, the searches carried out in Paper~I and this work very
clearly imply that the Be phenomenon is confined to a limited, if
broad, range of stellar parameters. High temperature stars do not
display the Be phenomenon, and high luminosity stars do not either.
If the results presented here are to be taken at face value, it may
even be thought that stars of $20-25M_{\sun}$ acquire Be
characteristics when they come into the 
range (in $T_{\rm eff}$ and $L_{*}$) of Be stars and then
lose them again as they move out towards high luminosities. Whether
this is telling us something important remains to be seen.

\section{Conclusions}
We have searched for new Oe stars by observing a sample of  early Be
stars and objects classified as Op and B0p. We find that Oe stars are
indeed very rare. Moreover almost all known Oe stars have
spectral types O9.5 or later, with only 4 stars known to have earlier
spectral types. The earliest spectral type is that of
\object{HD~155806}, O7.5\,III. The much higher fraction of Be stars
among B0-1 giants than among O8-9\,V stars, which are supposed to be
their progenitors, strongly argues for an increase of the
$\Omega/\Omega_{\rm crit}$ ratio during the MS lifetime of stars with
$M_{*}\approx20M_{\sun}$. Careful
determination of the stellar parameters of the few Oe stars known may
cast some light on the physical conditions leading to the Be phenomenon. 

\acknowledgements
The INT and WHT are
operated on the island of La 
Palma by the Isaac Newton Group in the Spanish Observatorio del Roque
de Los Muchachos of the Instituto de Astrof\'{\i}sica de
Canarias. Most of the observations presented here have been obtained
as part of the ING service programme and the authors are very thankful
to all the ING astronomers for their excellent work.

This research has made use
of the  Simbad data base, operated at CDS,
Strasbourg, France. This research has made use of the Washington
Double Star Catalog maintained at the U.S. Naval Observatory.

IN is a researcher of the
programme {\em Ram\'on y Cajal}, funded by the Spanish Ministerio de
Ciencia y Tecnolog\'{\i}a and the University of Alicante.
This research is partially supported by the Spanish Ministerio de
Ciencia y Tecnolog\'{\i}a under grant AYA2002-00814.

IN would like to thank Nolan Walborn for very careful reading of the
manuscript and many helpful comments and Artemio Herrero for very helpful
discussions. We also thank the referee, Thomas Rivinius, for his
helpful suggestions.

\end{document}